\documentclass[runningheads]{llncs}
\usepackage[T1]{fontenc}
\usepackage{graphicx}
\usepackage{algorithmic}
\usepackage[ruled,vlined]{algorithm2e}
\begin{document}
\title{CrossEAI: Using Explainable AI to generate better bounding boxes for Chest X-ray images}
\author{Jinze Zhao}
\authorrunning{Jinze Zhao}
\institute{University of Texas at Austin}
\maketitle   
% typeset the header of the contribution
%
\begin{abstract}
Explainability is critical for deep learning applications in healthcare which are mandated to provide interpretations to both patients and doctors according to legal regulations and responsibilities. Explainable AI methods, such as feature importance using integrated gradients, model approximation using LIME, or neuron activation and layer conductance to provide interpretations for certain health risk predictions. In medical imaging diagnosis, disease classification usually achieves high accuracy, but generated bounding boxes have much lower Intersection over Union (IoU). Different methods with self-supervised or semi-supervised learning strategies have been proposed, but few improvements have been identified for bounding box generation. Previous work shows that bounding boxes generated by these methods are usually larger than ground truth and contain major non-disease area. This paper utilizes the advantages of post-hoc AI explainable methods to generate bounding boxes for chest x-ray image diagnosis. In this work, we propose CrossEAI which combines heatmap and gradient map to generate more targeted bounding boxes. By using weighted average of Guided Backpropagation and Grad-CAM++, we are able to generate bounding boxes which are closer to the ground truth. We evaluate our model on a chest x-ray dataset. The performance has significant improvement over the state of the art model with the same setting, with $9\%$ improvement in average of all diseases over all IoU. Moreover, as a model that does not use any ground truth bounding box information for training, we achieve same performance in general as the model that uses $80\%$ of the ground truth bounding box information for training.

\end{abstract}
\section{Introduction}
Artiﬁcial intelligence (AI) is revolutionizing healthcare and, in particular, medical imaging. Health innovations applying machine learning (ML) and deep learning (DL) in radiology account for more than half of the total AI innovations in health. Advancing AI in medical imaging brings extraordinary beneﬁts with better accuracy, lower cost, and higher efﬁciency.  However, as disease classification in medical imaging achieves better performance by DL model, explainablity of the model, which is important for patients and doctors to trust the model, needs to be further explored. Chest X-rays are one of the most common images in non-invasive medical imaging diagnosis. In chest x-rays, generating bounding box of disease location is one way to achieve explainability. The bounding box generated is supposed to precisely includes the disease tissue, which is easy for doctors and patients to see the disease location and is convenient for radiologists to check the correctness. Previous work \cite{bib1} focused on using heatmap from DL model to generate bounding box, e.g. using Grad-CAM++ \cite{bib2}. However, the bounding box generated by this method is usually much larger than the ground truth bounding box and contain lots of non-disease areas. The low intersection over union (IoU) of ground truth and generated bounding box becomes the bottleneck for doctors to judge the exact location of the disease. In this work, we propose a new method that integrates two explainable AI methods to achieve better and focused bounding box with significantly higher IoU. We use weighted average of Guided Backpropagation \cite{bib3} and Grad-CAM++ \cite{bib2} to generate a new map of the chest x-ray image and draw a bounding box according to our bounding box generation algorithm. We test our model on a chest x-ray dataset \cite{bib4}, and our performance is significantly better than previous methods \cite{bib4,bib1} on all diseases. Compared with the state of the art with same setting, our result achieves significant margin with average $9\%$ improvement in average of all diseases over all IoU. Moreover, we also achieve same performance in general as the model that uses $80\%$ of the ground truth bounding box information for training, while our model does not use any ground truth bounding box information for training. As the bounding box generated by our model is smaller and closer to ground truth, and the gradient map generated by guided backpropagation can also highlight the exact disease location. Figure 1 shows an overview of our model.

\begin{figure}
\includegraphics[width=\textwidth]{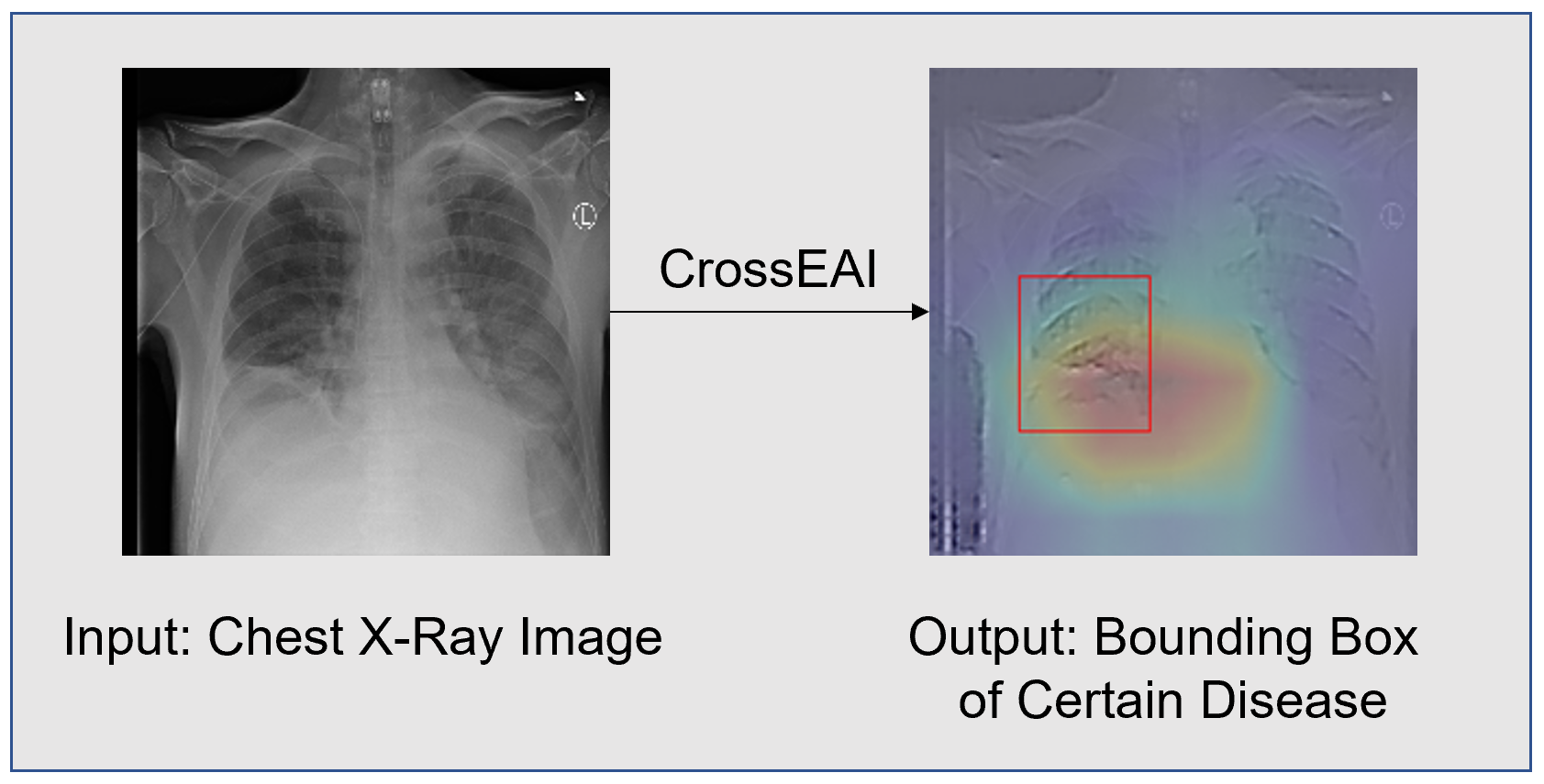}
\caption{An overview of our model. With an input of chest x-ray image, we output the weighted average of the gradient map and heatmap of the image to generate a bounding box of the disease location according to our bounding box generation method.} \label{fig1}
\end{figure}

In summary, this paper has the following contributions:
1) Significant improvement on bounding box generation: Our CrossEAI model has $9\%$ of improvement than the state-of-art model.
2) Cross validation using explainable AI: Interpretations can be subjective and cross validation of two explainable AI methods is valuable to identify the shared interpretations by both methods. In this paper, we use the weighted average to integrate weights from both explainable AI methods and propose the bounding box generation method based on the integrated weights to draw the bounding box on chest x-ray images. Previous methods generate much larger bounding boxes for chest x-rays, especially when IoU increases. Our CrossEAI has significant outperformed the baseline methods and generated better and more focused bounding boxes on NIH ChestX-ray8 dataset.
3) Disease details in chest x-ray image: We are the first to show details in the bounding box that gives disease outlines at the disease location. These details may help doctors better diagnose the disease.

\section{Related Work}
\subsection{Localization in Computer Vision}
Localization is one of the most important explainability in computer vision. It highlights the important pixels of the image, which contributes to the DL model to make the correct prediction. Guided Backpropagation \cite{bib3} and Deconvolution \cite{bib5} are two classic methods that use backpropagation to get the gradients of the input pixels. They have the advantage of high-resolution and they can highlight the fine-grained details in the image. However, they are not class-discriminative. When facing with multiple label prediction, this disadvantage becomes a big problem. Recently, heatmap as a class-discriminative method has been used a lot. \cite{bib6} proposed a Class Activation Mapping (CAM) method that uses global average pooling to indicate the discriminative image regions. Grad-CAM \cite{bib7} improved the performance by using the gradients flowing into the final convolutional layer to produce a coarse localization map highlighting the important regions of a image. Grad-CAM++ \cite{bib2} further improved the performance by using a weighted combination of the positive partial derivatives of the last convolutional layer feature and the specific class score. Although heatmap is class-discriminative, it does not have the advantage of high-resolution, which is obvious in medical imaging.

\subsection{Bounding Box Generation in Chest X-ray}
In chest x-ray, most previous work uses heatmap to generate bounding box. \cite{bib4} used LSE pooling \cite{bib8} to generate heatmap and found bounding box with ad-hoc thresholding. SCALP \cite{bib1} used Grad-CAM++ \cite{bib2} and dynamic programming to generate bounding box for the chest x-ray image. ChexRadiNet \cite{bib9} used radiomic features combined with CAM \cite{bib6} to generate bounding box for the chest x-ray image. However, with the use of heatmap, these models usually draw much larger bounding box than the ground truth which contain lots of non-disease areas. \cite{bib10} used patch slicing to generate more targeted bounding box. However, it needed ground truth bounding box supervision and used $80\%$ of the ground truth bounding box information in training.

\section{Method}
\subsection{Classification Model}
SCALP model \cite{bib1} is selected as our classification model, which uses patient-based supervised contrastive learning \cite{bib12} to improve the disease classification. The model structure is illustrated in Figure 2. ResNet-50 \cite{bib13} with pretrained weight provided by \cite{bib14} is the backbone of our classification model.
The contrastive learning module uses triplet attention \cite{bib11} that captures the interaction between the spatial and channel dimension of the input image. Patient metadata is utilized here as data augmentation for contrastive learning. For positive sampling, with a certain query image, we randomly select a chest X-ray image with the same disease label of the same patient using patient metadata as the positive sample. For negative sampling, we randomly select $k$ chest X-ray images with the same disease label of different patients using patient metadata as the negative samples. The contrastive loss is calculated by 
\begin{equation}
    L_{Contrastive} = -log \frac{\exp(sim(z_i,z_j)/\tau)}{\sum_k \exp(sim(z_i,z_k)/\tau)}
\end{equation}
where $z_i$ is the embedding of the quary image, $z_j$ is the embedding of the positive sample, $z_k$ is the embedding of a negative sample, $\tau$ is a temperature parameter and $sim(x,y)$ is the cosine similarity of $x$ and $y$.
The supervised classification module uses ResNet-50 to get a 2048-dimension feature vector from input chest x-ray image. A MLP layer is followed to predict the label of the chest x-ray image. For each label, we define a binary cross-entropy loss. The total cross-entropy loss is
\begin{equation}
    L_{Cross-Entropy} = \sum_n -y_n \cdot \log \hat{y}_n - (1-y_n) \cdot \log(1 - \hat{y}_n)
\end{equation}
where $y_n$ is the ground truth label of class $n$ and $\hat{y}_n$ is the predicted label of class $n$.
The total loss is calculated by the contrastive loss and cross-entropy loss together
\begin{equation}
    L_{Total} = \lambda \cdot L_{Cross-Entropy} + (1 - \lambda) \cdot L_{Contrastive}
\end{equation}
where $\lambda$ is a hyperparameter to control the proportion of contrastive loss and cross-entropy loss.

\begin{figure}
\includegraphics[width=\textwidth]{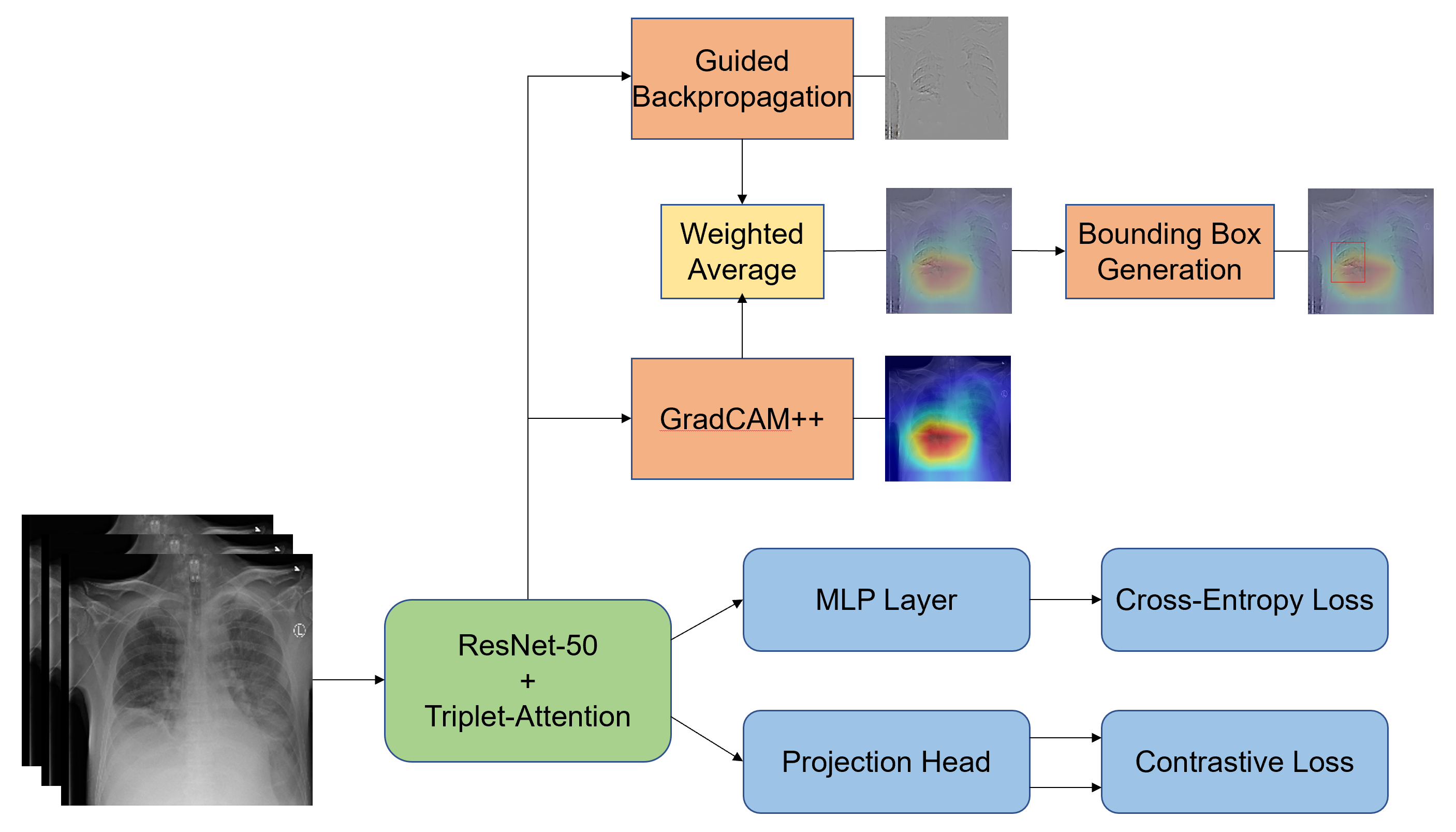}
\caption{Model Structure of CrossEAI. ResNet-50 with contrastive learning is first trained for classification. Then gradient map from Guided Backpropagation and heatmap from Grad-CAM++ are combined to generate the bounding box for the input chest x-ray image.} \label{fig2}
\end{figure}

\subsection{Localization Method}
\subsubsection{Heatmap Generation}
We use Grad-CAM++ \cite{bib2} to extract the $4^{th}$ layer heatmap for each chest x-ray image from our classification model. The heatmap generated has the advantage of class-discriminative, but usually contains lots of non-disease areas. As shown in Figure 3, the generated heatmap includes both disease location and non-disease areas. Although it contains major area of disease location, the more non-disease areas make doctors hard to find the true disease location and give a low IoU with ground truth.
\subsubsection{Gradient Map Generation}
We use Guided Backpropagation \cite{bib3} to extract the gradient for each chest x-ray image by doing backpropagation to the input layer of our classification model. The gradient map has the advantage of high-resolution and shows details, but it is not class-discriminative. As shown in Figure 3, the generated gradient map shows the outlines of the disease location, but it also highlights other non-disease area that may contribute to other disease in the classification part. Therefore, we cannot generate a bounding box only based on gradient map. Non-disease details need to be filtered to generate the true bounding box.
\subsubsection{Weighted Average of Gradient Map and Heatmap}
Heatmap and gradient map each has their advantage and disadvantage. To fully use the advantages of both maps and eliminate the disadvantages of each maps, we consider generating a new map by using weighted average to combine gradient map and heatmap, which is calculated by
\begin{equation}
    \textbf{New map} = t \cdot \textbf{Heatmap}  + (1-t) \cdot \textbf{Gradient map}
\end{equation}
where $t$ is a hyperparameter to control the proportion of gradient map and heatmap.
The new map is high-resolution due to contribution of gradient map. It is also class-discriminative because the low intensity areas in the heatmap, which is the non-disease areas, filter the high gradient from other class in the gradient map. Therefore, both advantages are kept and all the disadvantages are eliminated.
For bounding box generation, a pixel intensity threshold is then applied to filter the new map. After that, we use dynamic programming \cite{bib1} to generate a set of candidate rectangles and select the candidate which has the highest average intensity as our bounding box. Algorithm 1 describes our method in details.

\begin{algorithm}
\DontPrintSemicolon
  \textbf{Input:} Chest X-ray image\\
  \textbf{Output:} Coordinates (x1, y1, x2, y2) of the bounding box of certain disease \\
  \textbf{1.} Use Guided Backpropagation to generate gradient map from ResNet-50 \\
  \textbf{2.} Use Grad-CAM++ to generate $4^{th}$ layer attention map (heatmap) from ResNet-50 \\
  \textbf{3.} Scale gradient map intensities and heatmap intensities to [0, 255] \\
  \textbf{4.} Calculate weighted average of gradient map and heatmap to generate the new map\\
  \textbf{5.} Create a mask matrix with same dimension as the new map\\
  \textbf{6.} \If{pixel\_intensity > threshold}
    {
       mask[pixel] = 1
    }
    \Else
    {
        mask[pixel] = 0
    }
    \textbf{7.} Mask the new map with created mask \\
    \textbf{8.} Use dynamic programming to generate maximum area rectangles as candidate bounding box.\\
    \textbf{9.} Expand candidate rectangles uniformly across the edge until newly added \textit{ratio (0s count, 1s count)} > 1 \\
    \textbf{10.} Select the rectangle with the maximum average pixel intensity mapped in the masked new map and return its coordinates.
%   $\sum_{i=1}^{\infty} := 0$ \tcp*{this is a comment}
%   \tcc{Now this is an if...else conditional loop}
%   \If{Condition 1}
%     {
%         Do something    \tcp*{this is another comment}
%         \If{sub-Condition}
%         {Do a lot}
%     }
%     \ElseIf{Condition 2}
%     {
%     	Do Otherwise \;
%         \tcc{Now this is a for loop}
%         \For{sequence}    
%         { 
%         	loop instructions
%         }
%     }
%     \Else
%     {
%     	Do the rest
%     }
    
%     \tcc{Now this is a While loop}
%   \While{Condition}
%   {
%   		Do something\;
%   }

\caption{Bounding Box Generation Algorithm}
\end{algorithm}

\begin{figure}
\includegraphics[width=\textwidth]{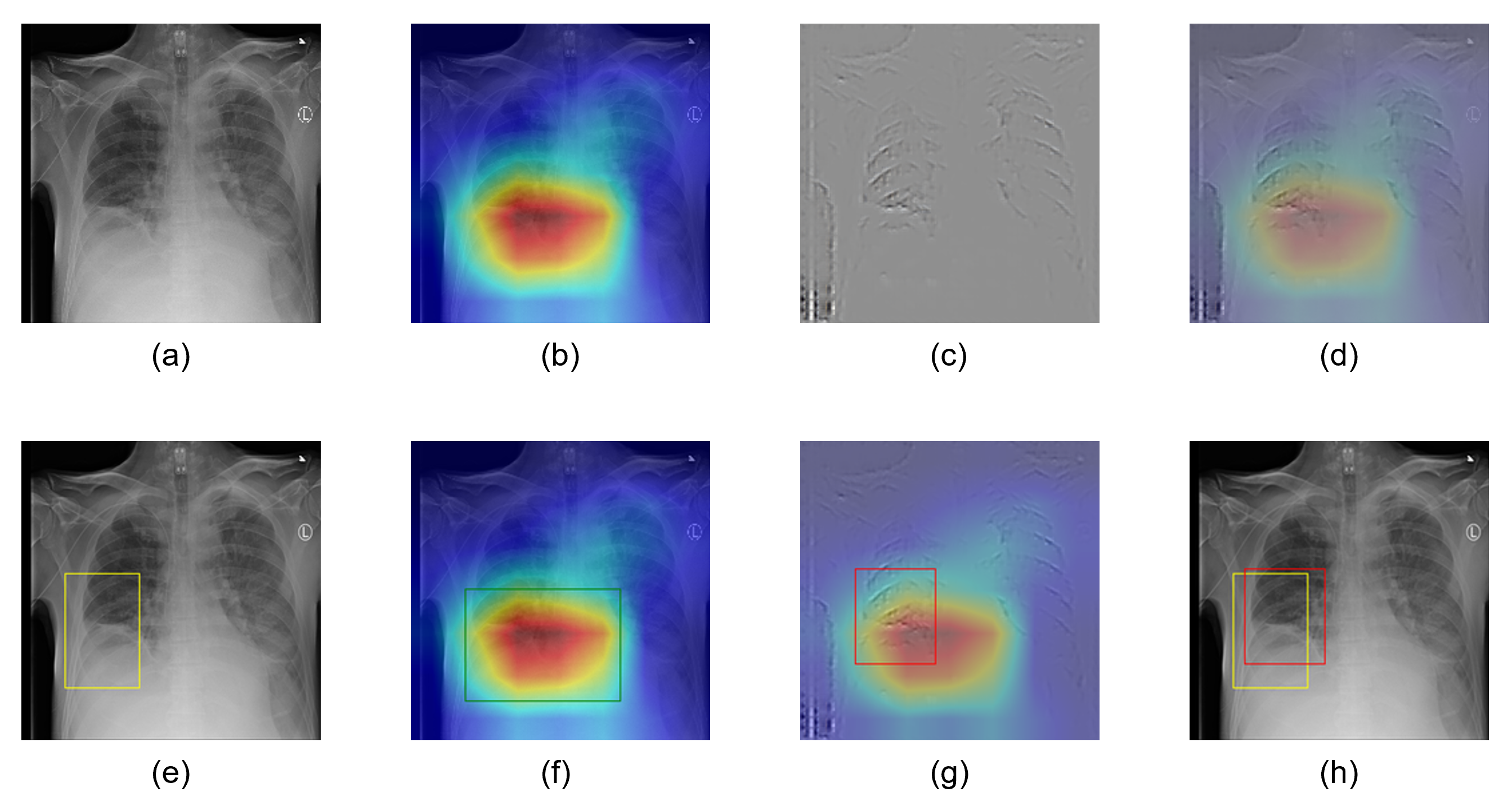}
\caption{An example of chest x-ray image with bounding box generated by different methods. (a) Original chest x-ray image; (b) Heatmap generated by Grad-CAM++; (c) Gradient map generated by Guided Backpropagation; (d) Weighted average of gradient map and heatmap; (e) Ground truth bounding box; (f) Bounding box generated only by heatmap; (g) Bounding box generated by our method; (h) Bounding box comparison of ground truth and our method.} \label{fig3}
\end{figure}

\section{Result}
\subsection{Dataset}
We evaluate our bounding box generation method on NIH ChestX-ray8 dataset \cite{bib4}. NIH ChestX-ray8 dataset consists of $112,120$ chest X-ray images collected from $30,805$ patients. It contains $8$ diseases, Atelectasis, Cardiomegaly, Effusion, Infiltration, Mass, Nodule, Pneumonia, and Pneumothorax. The labels are extracted from the associated radiology report by using an automatic labeler. We use these labels as ground truth. This dataset includes high-quality bounding box annotations for $880$ images by expert radiologists. We use our method to generate bounding box for each these images and compare them with the ground truth given by the radiologists.

\subsection{Implementation Details}
For the classification part, we use pretrained ResNet-50 model with triplet attention given by \cite{bib14}. The dataset is divided to three part, $70\%$ as training set, $10\%$ as validation set, and $20\%$ as test set. The batch size is $128$ and learning rate is $0.01$ with weight decay of $10^{-4}$ for cross-entropy loss and $10^{-6}$ for contrastive loss. $\lambda$ is set to be $0.80$.
For the bounding box generation part, $t$ is $0.30$. $30\%$ of the heatmap and $70\%$ of the gradient map are combined to form the new map. The pixel intensity threshold is set to be $35\%$ of the maximum pixel intensity.

\subsection{Experiment Results}

\begin{table*}
\centering
\begin{tabular}{clccccc}
\textbf{T(IoU)} & \textbf{Label} & \textbf{Baseline} & \textbf{Supervision} & \textbf{SCALP} & \textbf{ChexRadiNet} & \textbf{CrossEAI} \\
\hline
& Atelectasis & 0.69 & 0.71 & 0.62 & 0.72 & \textbf{0.73}\\
& Cardiomegaly & 0.94 & 0.98 & 0.97 & 0.96 & \textbf{1.00}\\
& Effusion & 0.66 & 0.87 & 0.64 & 0.81 & \textbf{0.87}\\
\textbf{0.1} & Infiltration & 0.71 & 0.92 & 0.81 & 0.88 & \textbf{0.92}\\
& Mass & 0.40 & \textbf{0.71} & 0.51 & 0.67 & 0.69\\
& Nodule & 0.14 & \textbf{0.40} & 0.12 & 0.33 & 0.37\\
% & Pneumonia & 0.63 & 0.80 & \textbf{0.93}\\
% & Pneumothorax & 0.38 & 0.29 & \textbf{0.83}\\
& Mean & 0.59 & \textbf{0.77} & 0.61 & 0.73 & 0.76\\
\hline
& Atelectasis & 0.47 & 0.53 & 0.42 & 0.49 & \textbf{0.53}\\
& Cardiomegaly & 0.68 & 0.97 & 0.92 & 0.84 & \textbf{0.99}\\
& Effusion & 0.45 & \textbf{0.76} & 0.42 & 0.62 & 0.73\\
\textbf{0.2} & Infiltration & 0.48 & \textbf{0.83} & 0.60 & 0.54 & 0.80\\   
& Mass & 0.26 & \textbf{0.59} & 0.25 & 0.46 & 0.49\\
& Nodule & 0.05 & \textbf{0.29} & 0.04 & 0.21 & 0.25\\
% & Pneumonia & 0.35 & 0.56 & \textbf{0.84}\\
% & Pneumothorax & 0.23 & 0.18 & \textbf{0.72}\\
& Mean & 0.40 & \textbf{0.66} & 0.44 & 0.53 & 0.63\\
\hline
& Atelectasis & 0.24 & 0.36 & 0.29 & 0.28 & \textbf{0.37}\\
& Cardiomegaly & 0.46 & 0.94 & 0.78 & 0.73 & \textbf{0.96}\\
& Effusion & 0.30 & 0.56 & 0.23 & 0.54 & \textbf{0.61}\\
\textbf{0.3} & Infiltration & 0.28 & \textbf{0.66} & 0.37 & 0.43 & 0.65\\
& Mass & 0.15 & \textbf{0.45} & 0.13 & 0.38 & 0.35\\
& Nodule & 0.04 & 0.17 & 0.01 & 0.15 & \textbf{0.18}\\
% & Pneumonia & 0.17 & 0.40 & \textbf{0.67}\\
% & Pneumothorax & 0.13 & 0.05 & \textbf{0.65}\\
& Mean & 0.25 & 0.52 & 0.30 & 0.42 & \textbf{0.52}\\
\hline
& Atelectasis & 0.09 & \textbf{0.25} & 0.18 & 0.17 & 0.23\\
& Cardiomegaly & 0.28 & \textbf{0.88} & 0.55 & 0.65 & 0.86\\
& Effusion & 0.20 & 0.37 & 0.12 & 0.42 & \textbf{0.46}\\
\textbf{0.4} & Infiltration & 0.12 & \textbf{0.50} & 0.19 & 0.32 & 0.48\\
& Mass & 0.07 & \textbf{0.33} & 0.09 & 0.29 & 0.28\\
& Nodule & 0.01 & \textbf{0.11} & 0.01 & 0.09 & 0.10\\
% & Pneumonia & 0.08 & 0.25 & \textbf{0.44}\\
% & Pneumothorax & 0.07 & 0.02 & \textbf{0.58}\\
& Mean & 0.13 & 0.41 & 0.19 & 0.32 & \textbf{0.42}\\
\hline
& Atelectasis & 0.05 & \textbf{0.14} & 0.07 & 0.11 & 0.12\\
& Cardiomegaly & 0.18 & \textbf{0.84} & 0.33 & 0.59 & 0.73\\
& Effusion & 0.11 & 0.22 & 0.04 & 0.29 & \textbf{0.37}\\
\textbf{0.5} & Infiltration & 0.07 & 0.30 & 0.10 & 0.15 & \textbf{0.36}\\
& Mass & 0.01 & \textbf{0.22} & 0.04 & 0.12 & 0.16\\
& Nodule & 0.01 & 0.07 & 0.00 & \textbf{0.07} & 0.05\\
% & Pneumonia & 0.03 & 0.14 & \textbf{0.32}\\
% & Pneumothorax & 0.03 & 0.10 & \textbf{0.50}\\
& Mean & 0.07 & 0.30 & 0.10 & 0.22 & \textbf{0.30}\\
\hline
& Atelectasis & 0.02 & 0.07 & 0.02 & 0.06 & \textbf{0.08}\\
& Cardiomegaly & 0.08 & \textbf{0.73} & 0.14 & 0.37 & 0.46\\
& Effusion & 0.05 & 0.15 & 0.02 & 0.09 & \textbf{0.29}\\
\textbf{0.6} & Infiltration & 0.02 & 0.18 & 0.04 & 0.06 & \textbf{0.23}\\
& Mass & 0.00 & \textbf{0.16} & 0.03 & 0.08 & 0.14\\
& Nodule & 0.01 & 0.03 & 0.00 & 0.04 & \textbf{0.03}\\
% & Pneumonia & 0.02 & 0.07 & \textbf{0.21}\\
% & Pneumothorax & 0.03 & 0.00 & \textbf{0.36}\\
& Mean & 0.03 & 0.22 & 0.04 & 0.12 & \textbf{0.25}\\
\hline
& Atelectasis & 0.01 & \textbf{0.04} & 0.01 & 0.02 & 0.03\\
& Cardiomegaly & 0.03 & \textbf{0.52} & 0.04 & 0.21 & 0.22\\
& Effusion & 0.02 & 0.07 & 0.01 & 0.04 & \textbf{0.19}\\
\textbf{0.7} & Infiltration & 0.00 & 0.09 & 0.03 & 0.02 & \textbf{0.15}\\
& Mass & 0.00 & 0.11 & 0.01 & 0.07 & \textbf{0.12}\\
& Nodule & 0.00 & 0.01 & 0.00 & 0.01 & \textbf{0.03}\\
% & Pneumonia & 0.01 & 0.02 & \textbf{0.13}\\
% & Pneumothorax & 0.02 & 0.00 & \textbf{0.28}\\
& Mean & 0.01 & \textbf{0.14} & 0.02 & 0.06 & 0.12\\
\hline
\end{tabular}
\vspace{0.3cm}
\caption{\label{table1}
Comparison of performance of different methods under different IoU. Note that Supervision \cite{bib10} uses $80\%$ images with ground truth bounding box for training. We do not use any ground truth bounding box information on training and get same performance with Supervision in average. Considering methods with same setting, we have $9\%$ improvement in average compared with ChexRadiNet \cite{bib9}.
}
\label{tab:result}
\vspace{-0.1cm}
\end{table*}

Figure 3 is an example of comparison between ground truth, bounding box generated only by heatmap, and bounding box generated by our model. (a)(b)(c) are the original chest x-ray image, heatmap generated from the input image, and gradient map generated from the input image separately. (d) is the weighted average of gradient map and heatmap generated by our model. As it shows, the heatmap part helps filter the non-disease details of the gradient map part. The gradient map part helps highlight the disease outline in the heatmap part. This combination utilizes both advantages of the gradient map and heatmap and gives a better map for bounding box generation. (e) shows the yellow ground truth bounding box of the input chest x-ray image. Compared with (e), the green bounding box in (f), which is generated only by heatmap, contains lots of non-disease area and therefore gives a low IoU. Red bounding box in (g) is generated by (d) using our model. (h) is a comparison of the ground truth bounding box and bounding box generated by our model. They are close to each other and therefore give a high IoU, which indicates the performance of our CrossEAI model. The details showed in (d) also give specific disease outlines which may help doctors to diagnose the disease.

Table 1 shows the result of different bounding box generation methods evaluated by different IoUs. We compare our performance with baseline \cite{bib4}, Method under supervision \cite{bib10}, SCALP \cite{bib1}, and ChexRadiNet \cite{bib9}. Our method continues getting better performance on all labels compared to methods with same setting, with $9\%$ improvement in average of all diseases over the state of the art heatmap method ChexRadiNet. We also achieve same performance in general as the method under supervision that uses $80\%$ of the ground truth bounding box information for training, which further demonstrate the performance of our model.

\bibliographystyle{splncs04}
\bibliography{mybibliography}

\end{document}